\documentclass[sigconf]{acmart}

\def\BibTeX{{\rm B\kern-.05em{\sc i\kern-.025em b}\kern-.08emT\kern-.1667em\lower.7https://www.overleaf.com/6667538993xdmcxnbtdbszex\hbox{E}\kern-.125emX}}
\usepackage{adjustbox}
\usepackage{makecell}

\definecolor{camera}{RGB}{65, 152, 182}

\newcommand{\etal}{\textit{et al. }}
\usepackage{soul}

\renewcommand\footnotetextcopyrightpermission[1]{} 
\begin{document}
\fancyhead{}
\title{The Role of Word-Eye-Fixations for Query Term Prediction}

%


\author{Masoud Davari, Daniel Hienert, Dagmar Kern, and Stefan Dietze}
\affiliation{%
  \institution{GESIS - Leibniz Institute for Social Science,\linebreak Cologne, Germany}}
  \email{{firstname.lastname}@gesis.org}

\begin{abstract}
Throughout the search process, the user's gaze on inspected SERPs and websites can reveal his or her search interests. Gaze behavior can be captured with eye tracking and described with word-eye-fixations. Word-eye-fixations contain the user's accumulated gaze fixation duration on each individual word of a web page. In this work, we analyze the role of word-eye-fixations for predicting query terms. We investigate the relationship between a range of in-session features, in particular, gaze data, with the query terms and train models for predicting query terms. We use a dataset of 50 search sessions obtained through a lab study in the social sciences domain. Using established machine learning models, we can predict query terms with comparably high accuracy, even with only little training data. Feature analysis shows that the categories \textit{Fixation}, \textit{Query Relevance} and \textit{Session Topic} contain the most effective features for our task.
\end{abstract}

%
%



%

%
\maketitle

\section{Introduction}
User's gaze behavior throughout a search session has been used in Information Retrieval (IR) as a source of implicit user and relevance feedback. It has been applied to understand the user interest \cite{ajanki2009can}, knowledge level \cite{cole2013inferring}, a viewed document's relevance \cite{puolamaki2005combining} or the overall task type \cite{liu2010search}. Resulting insights from gaze behavior has been used, for example, for re-ranking results or query expansion \cite{buscher2008query}. 

However, the examination of users' gaze behavior on the \textit{textual level} has been a hard and costly task so far, as standard eye tracking software only captures x,y-coordinates. The mapping from eye coordinates to actual text has to be done for each experiment from scratch. With the open source software \textit{Reading Protocol} \cite{HienertKMSB19} it is possible to automatically process all eye fixations on individual words of viewed web pages in a search session resulting in precise data about word-eye-fixations (duration, frequency, and timestamps).

While word-eye-fixations capture a part of the user's viewing and reading behavior, its role for different components of the IR search process is still insufficiently investigated. In this work, we address this issue by using a dataset of 50 search sessions to understand the role of word-eye-fixations and a range of other features for query term prediction. We compare several feature sets and build classification models for the task of query prediction, using established supervised classification approaches, such as Random Forest (RF), Logistic Regression (LR), Support Vector Machine (SVM), and Naive Bayes (NB). Thereby, we address the following research question: What feature configuration illustrates a better performance in predicting the future query terms in a session?

Our results show that, even given the small size of our dataset, the investigated features are effective in predicting future query terms and the feature categories \textit{Fixation}, \textit{Query Relevance} and \textit{Session Topic} contain the most important features.

\section{Related work}
In the following, we report on some works in the field of IR, which analyze reading behavior at the levels of paragraph, text and queries. 
Puolamaki \etal \cite{puolamaki2005combining} studied the user's gaze behavior while reading a list of titles from scientific articles. Gaze data in combination with trained Hidden Markov Model (HMM) could be used to predict the relevance for new document titles. Balatsoukas and Ruthven \cite{balatsoukas2010use} did a study on reading behavior on search engine result pages. They found that fixations were longer on relevant topically related surrogates of the SERP. On the paragraph level, Brooks \etal \cite{brooks2006measuring} found that relevant passages in a text have a higher number of fixations and regressions. Buscher \etal \cite{buscher2008query} used eye-gazed features, e.g. eye movements, fixations, and saccades to find relevant paragraphs. Ajanki \etal \cite{ajanki2009can} trained a SVM classifier based on eye-gaze features on short topical documents from Wikipedia which the user has marked as relevant or not. They could then automatically construct queries from eye movements where no learning data is available. Lately, Jacucci \etal  \cite{jacucci2019integrating} introduce a model for predicting document relevance in literature search with signals from EEG and eye- tracking. These neurophysiological features were calibrated by showing topics form the corpus and let the users select relevant keywords to the topic.

Hienert and Lusky \cite{hienert2017all} found that for domain-specific search a large part of used query terms has been seen before in the search session. Eickhoff \etal \cite{eickhoff2015eye} found that terms acquired in web search queries are fixated longer than non-query terms. They also show that there is a semantic relationship between reformulation terms and eye-fixated terms. In this paper, we extend their work by predicting future query terms using features such as \textit{Session Topic}, \textit{Lexical}, \textit{Term Context}, and \textit{Browsing} in addition to eye-fixation and semantic proximity features.

\section{Experimental Context}
\subsection{User Study} \label{sec:study}
For this experiment, we use data from a user study \cite{Kern:2019:LLU:3295750.3298950} within a digital library for social science literature containing documents in German and English. 30 participants in the field of social sciences took part. Five participants were later excluded due to bad eye tracking data. From the rest of the 25 researchers, 16 were female, 9 male, with age ranging from 23 to 45 years (mean: 28.6, SD: 4.12). 14 held a bachelor degree, 10 a master degree and one is a postdoctoral researcher. The participants performed two tasks. In one task, they were asked to search for and to bookmark relevant publications to a topic they are familiar with. And in the other task, they had to bookmark publications to an unfamiliar topic. The participants themselves chose the topics in both cases. As the focus of the user study was on highlights in abstracts, participants performed one task with and the other without highlights in a counter-balanced order. As there is no significant effect of the highlights on the average fixation duration on a word in the abstract \cite{Kern:2019:LLU:3295750.3298950}, we treat all data in the following as there were no differences in presenting the documents. The eye gazes were recorded through the remote eye tracking device SMI iView RED 250 using a sampling rate of 60Hz. Subjects looked at 2,344 web pages (SERPs and detailed views of records which contain metadata about a selected document such as title, author, and abstract) which resulted in ~2.6 million rows of eye tracking data. The average session length is about 24 minutes, with 6.43 queries on average in each session. The average number of search terms in a query is 2.44. After the experiments, participants provided information about the selected topics. On average, the overall topic description contains 4.4 terms.

\subsection{Building Word-Eye-Fixations}
Standard eye tracking software can capture the user's gaze on different stimuli such as web pages. One disadvantage is that stimuli are only stored as images and videos, making it difficult to assign gaze data to viewed and read texts by the user. The Reading Protocol software \cite{HienertKMSB19} uses the original web page instead of images so that the fixation duration for each word on a web page can be determined exactly. The main outcome is a JSON for each user and stimuli with all viewed words, fixation duration and counts, and timestamps for each fixation. For each user from the experiment described above, we processed the data in Reading Protocol to get the word-eye-fixations.
To get an impression of how intensively our participants have read the presented content, we calculated the average percentage of terms fixated on the detailed view of the records which is $33.97\%$ (SD: $16.55\%$). To prepare the data for the prediction task and to make it denser, we combined the word-eye-fixation per stimuli to word-eye-fixations per session based on the word stems. On average, word-eye-fixations then contain 781.23 fixated terms, i.e., seen terms, per session (SD: 378.95).

\begin{table*}[tbh!]
  \centering
  \caption{Extracted features for the prediction of query terms.}
  \label{tab:extractedFeature}
  \begin{adjustbox}{width=0.99\textwidth}
  \begin{tabular}{|c|c|c|l|c|}\hline
 \thead{Category} &  \thead{Notation} & \thead{Feature type} & \thead{Feature description}&\thead{Corr($f_i$, $Q$)}\\\hline
     & $dur_{\tau_i}$ & Continuous & Eye fixation duration of a user on a term $\tau_i$ & \textbf{0.078} \\
     & $f_{\tau_i}$ & Continuous & Total number of times a term $\tau_i$ is fixated in a session& \textbf{0.087}\\
     
    Fixation & $time\_f$ & Continuous & Timestamp of a fixated term seen for the first time in a session & -0.012\\
    & $time\_l$ & Continuous & Timestamp of a fixated term seen for the last time in a session&0.020\\
    & $time\_len$ & Continuous & Time span ($time_l-time_f$)& 0.050 \\\hline
    & $term\_len$ & Continuous & Length of a fixated term in session $s$ & -0.010\\
      Lexical & $pos\_tag$ & Categorical & Part-Of-Speech tags of a fixated term in a session e.g. VB, NN, JJ& 0.005\\\hline
      & $Max\_Cos(\tau_i, Q)$ & Continuous & Maximum cosine similarity of $\tau_i$ to either of query terms $Q$&0.074\\
      Query & $lch\_sim$ & Continuous &Leacock Chodorow similarity&0.061\\
      Relevance & $res\_sim$ & Continuous & Resnik similarity& \textbf{0.119}\\
      & $lin\_sim$ & Continuous & Lin similarity& \textbf{0.099}\\\hline
      Session Topic & $is\_topic$ & Boolean & Whether a fixated term $\tau_i$ belongs to the topic of a session (Section \ref{sec:topic}) &  \textbf{0.087}\\\hline
      Term & $viewed\_f$ & Categorical & Category of the source a term was first seen &- 0.009\\
      Context & $viewed\_l$ & Categorical & Category of the source a term was last seen & 0.072\\\hline
      & $SERP\_num$ & Continuous & Total number of SERPs seen prior to a term fixation&0.055\\
    Browsing  & $detail\_num$ & Continuous & Total number of detailed views of records seen prior to a term fixation & 0.050\\
      & $total\_num$ & Continuous & Total number of browsed pages in a session & 0.070\\\hline
  \end{tabular}
  \end{adjustbox}
\end{table*}

\section{Predicting Query Terms: Features \& Classification Models}
\subsection{Task}
The goal of our machine learning task is to predict query terms from gaze data, in particular from word-eye-fixations. We model this as a binary classification problem. The main research question is: How well can future query terms be predicted?\newline
To investigate our research question, we split the session into two equal parts and aim at predicting the query terms of the second part of the session using signals obtained through the first part only.

\subsection{Features} \label{sec:method}
This section provides an overview of our term-specific features and the motivation behind them. \newline

\textbf{1. Fixation.} 
Since previous studies have shown that term-related fixation behaviour provides signals about term importance (e.g. \cite{eickhoff2015eye}) for query term acquisition, we consider fixation duration $dur_{\tau_i}$ and the fixation count $f_{\tau_i}$ as gaze-related measures which are extracted from our eye tracking data, where $\tau_i \in T$ and $T$ is set of all fixated terms.\newline
\textbf{2. Lexical.} This category contains the lexical features of the fixated word, e.g., term length and Part$-$Of$-$Speech (POS) tagging. In \cite{wang2016part} the effectiveness of leveraging POS tagging in IR tasks is shown.\newline
\textbf{3. Query Relevance.} Here, we calculate the maximum cosine similarity between a fixated term $\tau_i$ to either of query terms in $Q$ in each session $s$. To calculate the $Cos(\tau_i, Q)$, we used a pre-trained model of Word2Vec word embedding on German Wikipedia\footnote{https://github.com/devmount/GermanWordEmbeddings}. This model learns the word vectors with 300 features (dimensions) with a sliding window size $W_s = 5$. To measure other semantic proximity e.g., Leacock Chodorow \cite{leacock1998combining}, Lin \cite{lin1998information} and Resnik similarity \cite{resnik1995using}, we rely on GermaNet using its Python implementation\footnote{https://pypi.org/project/pygermanet/}. Computing semantic relatedness e.g., lch\_similarity is motivated by \cite{eickhoff2015eye}, and to the best of our knowledge is not yet used in other prediction tasks. \newline
\textbf{4. Session Topic.}  Here, we compute the topical term feature \newline($is\_topic$) in a session as explained in Section \ref{sec:topic}. \newline
\textbf{5. Term Context.} In \cite{hienert2017all}, it has been shown that the context of a fixated term is essential for its importance for the remaining session. For instance, whether it appears on a particular part of the SERP or detailed view of a record. 
Therefore, we consider the first/last context a term has been fixated on as features. \newline
\textbf{6. Browsing.} This category covers users' engagement with the system. We assume that it is more promising to predict query terms for a user who used the system more intensively. Therefore we consider the number of viewed SERPs and viewed detailed views of records as features. \newline
Table \ref{tab:extractedFeature} shows the features introduced above. Each fixated term is represented by a feature vector $\vec{f} = (f_1,f_2,f_3,f_4, ..., f_k)$. In the last column, the Pearson correlation coefficient of each feature with $Q$ is shown.

\subsection{Relation between Terms and Session Topics}\label{sec:topic}
For computing the \textit{is\_topic} feature, we make use of the thesaurus of the social science \textit{TheSoz}\footnote{http://lod.gesis.org/thesoz/en.html} and evaluate the results using the information about the actual topics provided by the participants. First, we extracted the topic concept and then computed the term-topic relevance. 

\textbf{Extracting topic concepts.} The list of word-eye-fixations contains all fixations on words which has been fixated on any SERPs and detailed views of records throughout the search session. As the words come from titles, abstracts, and other metadata they can be quite diverse. 
We use \textit{TheSoz} to disambiguate diverse terms to a controlled vocabulary. \textit{TheSoz} contains about 12,000 entries with 8,000 descriptors and 4,000 synonyms. First, we sort word-eye-fixations by fixation duration and count. We implemented an annotation algorithm\footnote{https://git.gesis.org/davarimd/parsingNT} to obtain a broader concept from the thesaurus for each fixated term  based on lexical similarity.
We use a Levensthein threshold of 3 based on our inspection on lexical similarity of the fixated term and the result list in \text{TheSoz}.
This way, we were able to add concepts to more than 78.23\% of the fixated terms with a fixation duration higher than 350ms, which is the mean fixation duration of fixated terms. On average, about 302.2 concepts are assigned to fixated terms.\newline

\textbf{Computation of term-topic relevance.}
To compute the relevance of a fixated term to a given session topic, we aggregate the fixation duration of all the fixated terms having the same concept. Then, we rank all concepts in a descending order according to their fixation duration. From this list we took the top 5 concepts and treat them as session topics. We evaluate our approach by calculating the average match of topics expressed by the participants as explained in section \ref{sec:study} and the automatically extracted topics. On average 72.44\% of the topic terms expressed by the participants can be found in the fixated terms of the top 5 concepts. We chose the top 5 concepts for our prediction task as this has the highest Pearson correlation with query terms ($r$ = 0.273) compared to top 10 ($r$ = 0.251), top 15 ($r$ = 0.226), top 20 ($r$ = 0.205) and top 25 ($r$ = 0.193).

\subsection{Classification models} \label{sec:classifiers}
We use standard supervised models for classification, namely Random Forest (RF), Logistic Regression (LR), Support Vector Machine (SVM), and Naive Bayes (NB) and use only features with positive Pearson correlation coefficient. For the experiment we used the scikit-learn library for Python.

\section{Experiments \& results} \label{sec:experiments}
\subsection{Experimental setup}

\textbf{Baseline.} 

In order to compare our model, we chose the Gaze-Length-Filter approach introduced in \cite{buscher2008query} as our baseline (\textit{RF-GLF}). In this method, Buscher et al. expand the query by computing the traditional \textit{tf-idf}  by using the gazed words in a passage and the number of fixations in those segments. For \textit{RF-GLF}, we compute \textit{tf-idf} for the fixated terms and the frequency of fixations.  As a text corpus, we take all visited documents by all participants in our eye-tracking study.

\textbf{Model variants.} 
To explore the influence of the different feature categories on model performance, we also investigate two model variants: (1) Random Forest using only query relevance features (\textit{RF-QR}) and (2) Random Forest using all features except fixation (\textit{RF-nF}). We chose Random Forest for this investigation as it is the best performing classifier in our experiment (see Table 2).

\textbf{Training and testing data.} 
The dataset consists of 50 sessions with an average of 781.23 unique fixated terms per session and the extracted features described above. As we model the prediction tasks as a binary classification problem, we concatenate the 50 sessions in a dataset which in total contains 26,187 fixated terms. For our task, query terms are annotated as class-1 in each session. To address our research question accurately, we exclude query terms which were used in both split parts of each session. That way, we predict the acquisition of query terms and not their recurrence. The datasets in the task was split into a training and a test set with 80\% of the dataset for training and 20\% of the dataset for validation. For all classifiers, we run 10-fold-cross-validation. 

\textbf{Class distribution and balancing.} The class distribution in our dataset is highly imbalanced, e.g., in our dataset there are 26,061 instances corresponding to class-0 and 126 instances corresponding to class-1. 

In order to prevent classifiers biased towards the majority class, we compare the performance of the classifier using under-sampling. 

\subsection{Results} \label{sec: result}

For the evaluation of the model, we use standard information retrieval metrics (Precision, Recall, and F\_1 score) and their macro average. Table \ref{tab:result} shows the performance of different configurations and baselines. The best performing model is Random Forest with $0.704$.
To analyze the influence of different features, we show the Pearson correlation between feature and query term in Table \ref{tab:extractedFeature} in the most right column. According to their correlation, the most effective features are related to \textit{Query Relevance} where \textit{res\_sim} has the highest correlation (0.119), followed by \textit{lin\_sim} (0.099). The next best features come from the \textit{Fixation} category with \textit{fixation count} (0.087) and \textit{fixation duration} (0.078). The category \textit{Session Topic} with \textit{is\_topic} showed a similar correlation of (0.087).

 \begin{table}[ht]
\caption{Performance (macro-average) of Run-time (s), Precision (P), Recall (R) and F1 for query term prediction using different classifiers and configuration. * marks the baseline.}
\label{tab:result}
\centering
\small
\begin{tabular}{|c|c|c|c|c|c|l|l|l|l|l|}\hline
Method & \thead{Run-time (s)} & \thead{Precision} & \thead{Recall}& \thead{F1}\\\hline
\thead{RF-GLF*} & 0.550 & 0.631 & 0.630 & 0.630\\\hline
\thead{RF-QR} &  0.684 & \thead{0.794} & \thead{0.762} & \thead{0.757}  \\\hline
\thead{RF-nF} & 0.565 & 0.650 & 0.645 & 0.644 \\\hline
\thead{RF} & 0.670 & \thead{0.708} & \thead{0.705} & \thead{0.704}\\\hline
\thead{LR} & 0.697 & 0.753 & 0.702 & 0.689 \\\hline
\thead{SVM} & 0.625 & 0.523 & 0.515 & 0.470 \\\hline
\thead{NB} & 0.697 & 0.724 & 0.662 & 0.640 \\\hline
\end{tabular}
\end{table}

\section{Discussion and Conclusion}
For the query term prediction tasks, we compared different classification models such as RF, LR, SVM, and NB. Random forest performed best with an $F_1$ score of $0.704$. Given the small size of the undersampled and imbalanced dataset, where only 252 instances were available in the experiment, these appear to be promising prediction results. This suggest that the investigated features are effective in predicting query terms. All models except SVM performed better than the baseline RF-GLF presented in \cite{buscher2008query}.

The most important features are \textit{res\_sim}, \textit{lin\_sim} from the group of query relevance, \textit{fixation count} and \textit{fixation duration} from the group of fixations and \textit{is\_topic} from the group session topic. Query relevance models the semantic proximity from queries to fixations. Session topic models the semantic proximity form the fixations to the overall session topics. Both features consider the semantic proximity on different levels, and represent the topics the user is searching for in the session. This seems to be important features for predicting future query terms. However, basic fixation measures such as fixation duration and fixation count also seems to be reasonable good features. 

The higher performance of RF-QR and the importance of the feature group in predicting future queries suggests that query relevance plays a major role in predicting queries, which is kind of intuitive.
We find that semantically similar terms to query terms are a good indicator for predicting future queries in a digital library which is similar to the findings from \cite{eickhoff2015eye} in Web search. Yet, the performance of RF-QR for predicting future queries, shows that with only using semantic similarity of fixated terms to query terms the performance would be better than using features from both groups \textit{Fixation} and \textit{Query Relevance}.

In addition to query relevance, we proposed the new feature \textit{is\_topic} which describes the fixation terms' semantic proximity to the overall session topic. One problem in using word-eye-fixations for prediction and other tasks is that they are distributed over different web pages, in different text passages and in different declension forms. With the presented method in Section 4.3, we are able to cluster fixations semantically to the broadest concept found in the controlled vocabulary of the thesaurus. With that, we can sum up fixations times and counts which belong to the same concept. This gives a much denser dataset of fixations and shows which concepts have been fixated by the user over the whole session. 

The model variant \textit{RF-nF} in which we used all features of query relevance and session topic but not the basic fixation measures performed comparably good (F1: 0.644 compared to F1 of RF with 0.704), which indicates that these derivative features work quite well for predicting future queries. Our goal, in future work, is to derivative features that can substitute the eye gaze data.
      
While these experiments are not yet aimed at predicting query term performance as such, our results suggest that the investigated features may be used for implementing effective query term recommenders. As part of future work, we are planning to investigate prediction performance on larger datasets, as well as additional features which we are currently inferring from eye tracking data, in particular concerning to their effectiveness for predicting query terms and their performance in search tasks.  


%

\bibliographystyle{ACM-Reference-Format}
\bibliography{biblio}
%

\end{document}